\documentclass{ws-procs9x6}

\def\be{\begin{equation}}
\def\ee{\end{equation}}  
\def\bea{\begin{eqnarray}}
\def\eea{\end{eqnarray}}

\def\la{\langle}
\def\ra{\rangle}

\begin{document}

\title{The scalar-isoscalar spectral function of strong matter
in a large N approximation}

\author{A. Patk{\'o}s}

\address{Department of Atomic Physics, E{\"o}tv{\"o}s University,
 H-1117 Budapest, Hungary}

\author{Zs. Sz{\'e}p}
\address{Research Group for Statistical Physics of the Hungarian Academy of
Sciences, H-1117, Budapest, Hungary}

\author{P. Sz{\'e}pfalusy}
\address{ Research Institute for Solid State Physics and Optics,\\
Hungarian Academy of Sciences, H-1525 Budapest, Hungary\\
\vspace*{0.1cm}
Department of Physics of Complex Systems,\\ 
E{\"o}tv{\"o}s University,
H-1117 Budapest, Hungary}

\maketitle


\abstracts{
The enhancement of the scalar-isoscalar spectral function near the two-pion
threshold is studied in the framework of an effective linear $\sigma$ model,
using a large N approximation in the number of the Goldstone bosons. The
effect is rather insensitive to the detailed $T=0$ characteristics of the 
$\sigma$ pole, it is accounted by a pole moving with increasing $T$
along the real axis of the second Riemann sheet towards the threshold
location from below.}

\section{Introduction}
With the advent of the new heavy-ion experiments, the significance of 
the theoretical study of models at high temperature and density is growing. 
The study started by Hatsuda {\it et al.} \cite{Hatsuda1}
of the $\sigma$ particle has recently received particular interest. 
At nonzero temperature $T$ and baryon density $n_B$ they conjectured 
the narrowing of the broad $\sigma$  resonance: during the restoration of
chiral symmetry, the mass of the $\sigma$ meson decreases which entails the
squeezing of the phase space available for the $\sigma\rightarrow 2\pi$ decay.
The gradual enhancement of the spectral function near the two-pion
threshold means that in spite of the elusive nature of the $\sigma$ particle 
in the vacuum, at $T=0$, $n_B=0$, there is a chance to see it more clearly in 
medium. 

Our aim is to study in the leading order large N limit of the linear
$\sigma$ model the trajectory of the sigma pole and its imprint on the 
shape of the scalar-isoscalar spectral function. Here, we point out only the 
main features of the investigation, referring for a more comprehensive 
treatment to Ref.~\cite{Patkos1} for the chiral limit and Ref. \cite{Patkos2} 
for the physical value of the pion mass.

\section{The model and the quantities of interest}

The Lagrangian of the linear $\sigma$ model including an
explicit symmetry breaking external field $h$ is parameterised 
as:\\
\centerline{$
\displaystyle L={1 \over 2}[(\partial\vec\phi)^2-
m^2\vec\phi^2]-{\lambda\over 24N} [\vec\phi^2]^2+\sqrt{N}h\phi_1.
$}
\\
In the broken symmetry phase one separates the vacuum expectation value of
the $\sigma$ field by the replacement:
$\phi^a\rightarrow(\sqrt{N}\Phi(T)+\phi^1,\phi^i)$.
For the $\sigma-\pi$ system, $N=4$ and one has $\Phi(0)=f_\pi/2$ where
$f_\pi=93$ MeV. The role of the external field $h$ is to give a mass to the 
pions: $m_\pi^2(T)=h/\Phi(T)$.

There are several reasons that favour the use of a large N
approximation. First of all since the expansion goes in powers of $1/N$, 
a strongly self-coupled theory does not pose any limitation to this 
expansion. This approach is insensitive to the
choice of the renormalisation point. It leads to a 
2$^\textrm{\small nd}$ order chiral transition and provides correct
critical description near $T_c$. It bears also some nice
features required in phenomenology: satisfies at LO
the  scalar-isoscalar channel unitarity and the  Adler-zero condition.

The quantities of interest in our investigation are the equation of state
$\la\phi^1\ra=0$, the $\sigma$ propagator $G_\sigma(p)$,  whose pole trajectory 
we are interested in, and its spectral function: 
$\rho_\sigma (p_0,T)=-{1\over\pi}
\lim\limits_{\varepsilon\rightarrow +0}{\rm Im}
G_\sigma(p_0+i\varepsilon,T)$.\\
Using the equation of state 
$
m^2+{\lambda \over 6}\Phi^2(T)+{\lambda\over 6N}
\la\phi^a\phi^a\ra
={h\over\Phi(T)}
\label{eqstate}
$
in the pion propagator, we check the
general Goldstone theorem for pions, {\it i.e.} their propagator
turns out to be $G_\pi(p)=1/(p^2-m_\pi^2(T))$. This propagator is used 
when calculating the pion bubbles both in the equation of state and in
the $\sigma$ propagator.

At leading order in the $1/N$ expansion $G_\sigma^{-1}(p)$ is
given by\\
$G_\sigma^{-1}(p)=p^2-m^2-
[\raisebox{-0.42cm}{\includegraphics[width=7.5cm]{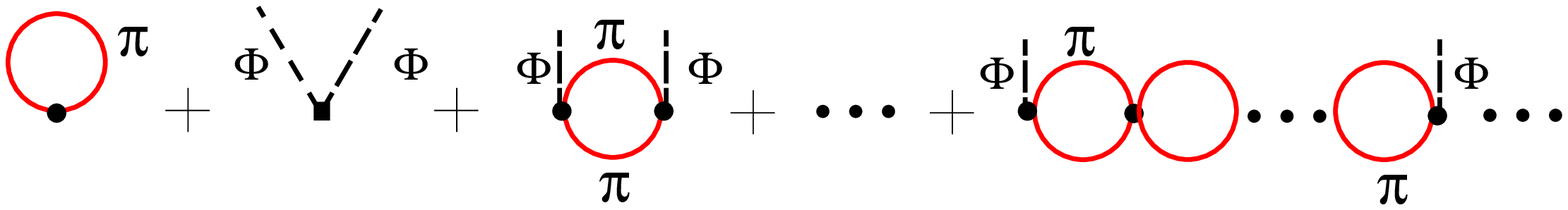}}].
$\\
With the help of the equation of state,  after performing a 
coupling constant renormalisation 
${1\over\lambda} +{1\over 96\pi^2}\ln{e\Lambda^2\over M_0^2}=
{1\over\lambda_R}$, one can express $G_\sigma^{-1}(p)$ as
\\ 
\vspace*{-0.3cm}
\centerline{$
G_\sigma^{-1}(p)=
p^2-\frac{h}{\Phi(T)}-\frac{\lambda_R\Phi^2(T)/3}{1-\lambda_R b_R(p)/3},
\qquad\qquad \textrm{where}\quad
b(p)=\raisebox{-0.425cm}{\includegraphics[width=.6cm]{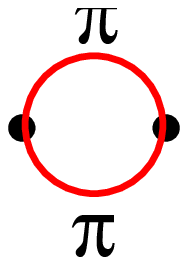}}. 
\label{Eq:sigma-prop}$}
\vspace*{-0.3cm}\\
\noindent
Notice, that the imaginary part of $G_\sigma$ is due exclusively to the
imaginary pieces in $b_R(p)$ ($b_R(p)$ is the finite part of $b(p)$).
In what follows we restrict ourselves to the propagator
with zero spatial  momentum $({\bf p}=0)$.

\section{Analytical continuation}

Since we are looking for poles of $G_\sigma(p_0)$ on the unphysical,
second Riemann sheet, where usually resonances lie, we have to perform the
analytical continuation of  the expression of $G_\sigma(p_0)$. 
This actually means the continuation of the function $b_R(p_0)$, which 
initially is defined on the physical sheet. We have chosen to perform the 
continuation of $b(p_0)$ across the positive real axis where we have to deal 
with the discontinuity along the real axis of the original expression for
the temperature dependent part of the bubble, $b_T(p_0)$ \cite{Patkos2}. 


The mass $M_\sigma$ and the width $\Gamma$ of the $\sigma$ particle are the 
real and the imaginary parts of the solution of $G^{-1}_\sigma(p_0)=0$, 
respectively. To determine the value of the coupling constant $\lambda_R$, 
we use the phenomenological value of $M_\sigma/\Gamma$ ratio at $T=0$ and 
the restriction imposed by the existence of a tachyonic pole in the theory. 
The tachyon, which is a pole of $G_\sigma(p_0)$ on the imaginary axis of the 
physical sheet, is intimately related to the effective nature of the 
$\lambda\Phi^4$ theory. By looking at the coupling renormalisation formula we 
see that, if one wants to evade triviality  ($\lambda_R=0$ ) maintaining the 
stability of the theory ({\it i.e.} $\lambda>0$) one has to keep the value of 
the cutoff finite. In order to have a stable effective theory the cutoff has to 
be below the scale of the tachyon. 

The value of the ratio $M_\sigma/\Gamma$ approaches the phenomenologically 
preferred value with increasing $\lambda_R$, but in the same time the tachyon 
mass comes closer to $M_\sigma$. Remaining  below the tachyonic scale by a 
factor of 3 to 5, the closest value is achieved for
$\lambda_R=310$ ($M_\sigma=3.5f_\pi$, $M_\sigma/\Gamma\sim1$) 
in the limiting chiral case and $\lambda_R=400$ ($M_\sigma=3.95f_\pi$,
$M_\sigma/\Gamma=1.4$) for $m_\pi(0)=140$ MeV.



\section{Results}

\begin{figure}[t]
\centerline{  
\includegraphics[width=6cm]{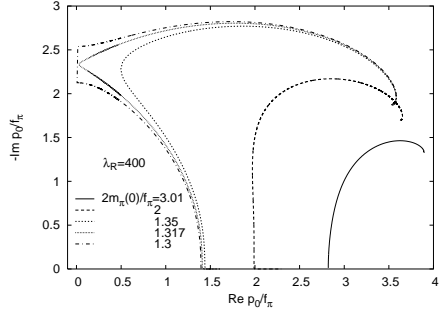}
}
\caption{Trajectory of the complex $\sigma$ pole for various values of 
$2m_\pi(0)/f_\pi$ in the 4$^\textrm{\scriptsize th}$ quadrant of the energy plane.}
\label{Fig:atcsapas}
\end{figure}

Fig.~\ref{Fig:atcsapas} shows the basic features of the temperature driven
$\sigma$ pole trajectory for different values of the $T=0$ pion mass. The
imaginary part of the pole will eventually decrease with
increasing $T$ and the pole approaches the two-pion threshold. But in a first 
stage the imaginary part actually increases, a feature observed 
also in the context of chiral perturbation theory \cite{Dobado}.

One can also see that with decreasing $m_\pi(0)$ the pole approaches more and 
more the imaginary axis. 
Below a certain mass value 
actually two oppositely moving poles emerge  from the collision on the
imaginary axis of the pole in the 4$^{\textrm{\small th}}$ quadrant with its 
3$^\textrm{\small rd}$ quadrant mirror. 
The one that moves towards the origin comes down from the imaginary axis 
switching over to the real axis only when colliding with one pole belonging
to the infinite chain of poles which are on the imaginary axis due to
the Bose-Einstein distribution function. In the chiral limit no second collision occurs, the 
pole goes to the origin as $T\rightarrow T_c$ along the negative imaginary
axis accounting for the observed critical behaviour~\cite{Patkos1}.
 
\begin{figure}[t]
\centerline{  
\includegraphics[width=6cm]{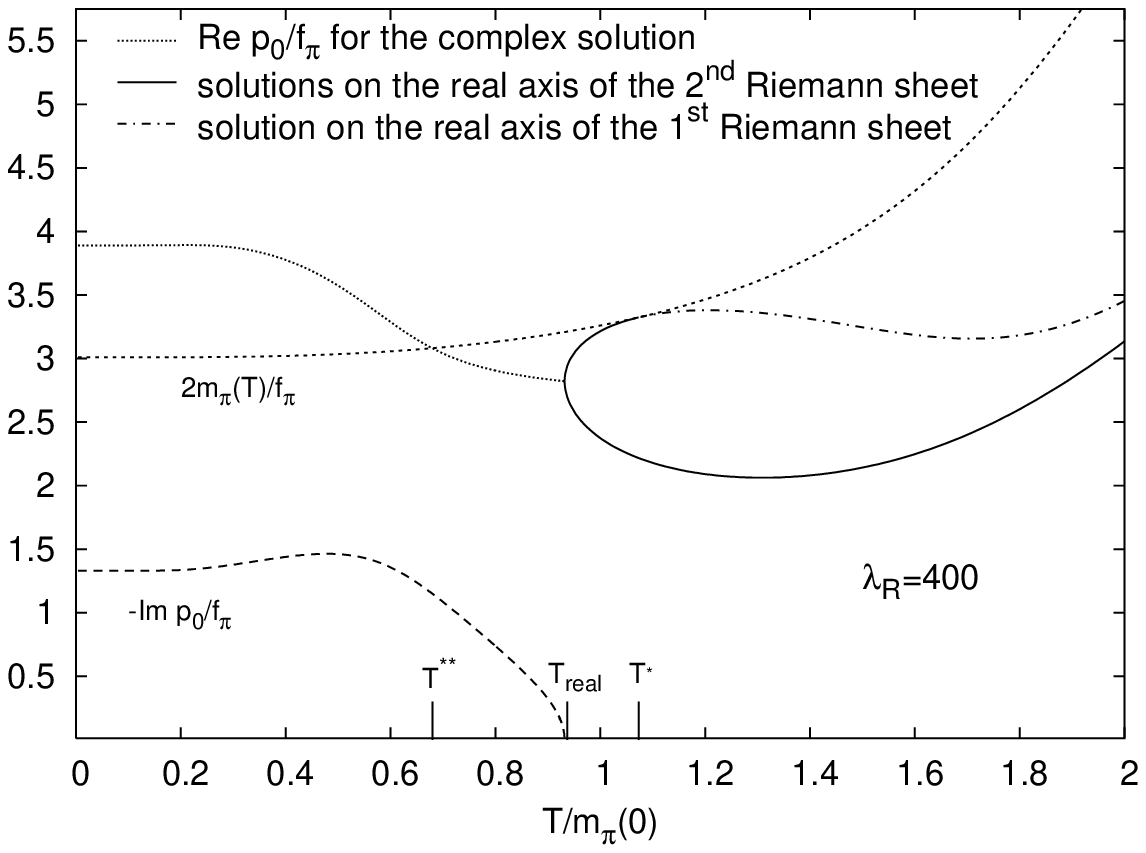}
\includegraphics[width=6cm]{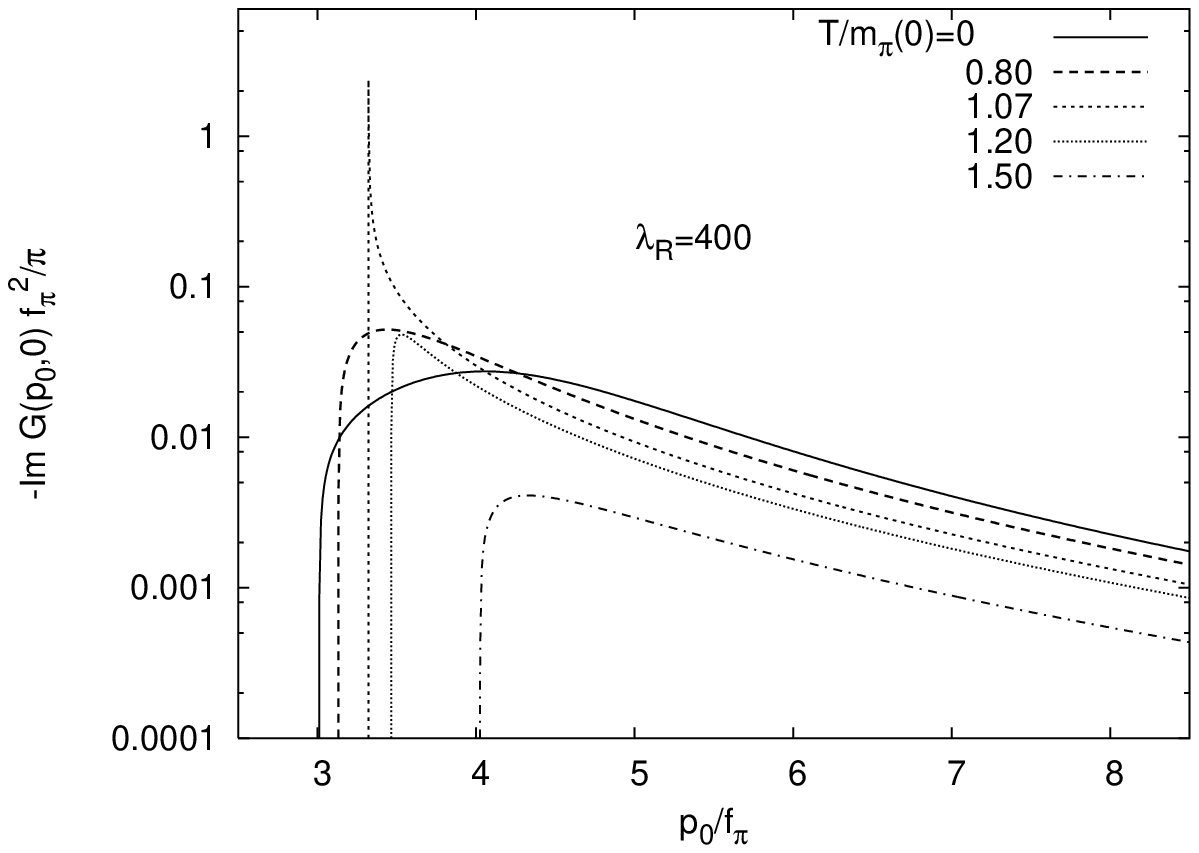}
}
\caption{The temperature dependent trajectory of the $\sigma$ pole for 
$m_\pi(0)=140$ MeV (left) and the variation of the spectral function
$\rho_\sigma$ with $T/m_\pi(0)$ (right).} 
\label{Fig:phys_pion}
\end{figure}

The l.h.s of Fig.~\ref{Fig:phys_pion} presents in more details the trajectory 
of the $\sigma$ pole for the physical pion mass $m_\pi(0)=140$ MeV.
At $T^{**}\approx 0.69 m_\pi(0)$ the real part of the pole 
goes {\it below} the threshold while the imaginary part is still finite.
Then at $T_\textrm{\small real}$ the pole collides with its
1$^\textrm{\small st}$
quadrant mirror on the real axis of the
2$^\textrm{\small nd}$ Riemann sheet,  below the 
threshold and splits up in two poles moving along the real axis.
At $T^{*}\approx 1.07 m_\pi(0)$ one of the poles reaches the   
threshold, climbs up to the 1$^{\textrm{\small st}}$ Riemann sheet and becomes 
physical, the other remains on the real axis of the unphysical sheet.
A pole moving on the real axis of the unphysical sheet towards the threshold
was recently reported
in Ref. \cite{Morimatsu} using an alternative approach to 
the $\sigma$ model. 

We observe by looking at the spectral function presented on the r.h.s of
Fig.~\ref{Fig:phys_pion} that only for $T<T^{**}$ there is trace of the 
$\sigma$ pole in the spectral function. For higher temperature the
approximate Lorentzian shape is distorted and in this way no resonance can
be identified. A gradual threshold enhancement occurs in the interval $T\in
(T^{**}, T^*)$, where the real part of the pole is smaller than $2m_\pi(T)$.
The maximum of the enhancement develops for $T=T^{*}$, around which  
$\rho_\sigma(p_0,T^*)\approx(1-4m_\pi^2(T^*)/p_0^2)^{-1/2}$. The details of
how this asymptotic behaviour sets in depend crucially on the fact that the
pole approaches the threshold along the real axis.

To test the generic nature of pole evolution presented above,
we have investigated at $T=0$ the dependence of the pole trajectory on the
baryon charge density $n_B$. Following Ref. \cite{Hatsuda2} the decrease of the 
chiral order parameter  with increasing values of $n_B$ in matter was
taken into account by linearly rescaling the $T=0$ vacuum expectation value. 
Qualitatively the same pattern of the pole trajectory was obtained as $n_B$
varies, like the one presented in Fig.~\ref{Fig:phys_pion}. This shows that
the scenario  is independent of the specific type of
thermodynamical driving force applied to the system.

\section{Conclusions}

In the LO large N approximation of the $\pi-\sigma$ system we presented a
generic scenario for the $\sigma$ pole evolution.
For a physical value of $m_\pi(0)$ the pole hits the real axis of the
unphysical sheet below 
the threshold and produces the threshold enhancement in the spectral function 
when moving on the real axis towards the threshold. This represents a 
qualitative prediction
for heavy ion collision experiments since the shape of the spectral
function in a wide temperature range 
would lead to  a high intensity
narrow peak in the $2\gamma$ spectra coming from the 
$\sigma\rightarrow 2\pi\rightarrow 4\gamma$ decay chain.

\end{document}